\documentclass[twocolumn,showpacs,prl,superscriptaddress]{revtex4}
\usepackage{graphicx}
\usepackage{color}
\usepackage{amssymb,amsfonts,amsmath}
\usepackage{natbib}

\begin{document}

\title{The second will be first: competition on directed networks}

\author{Giulia Cencetti} 
\affiliation{Dipartimento di Sistemi e Informatica, Universita` di Firenze,
Via S. Marta 3, IT-50139 Florence, Italy},
\author{Franco Bagnoli}  \affiliation{Universit\`{a} degli Studi di Firenze, Dipartimento di Fisica e Astronomia, CSDC and INFN, via G. Sansone 1, 50019 Sesto Fiorentino, Italia}
\author{Francesca Di Patti} \affiliation{Universit\`{a} degli Studi di Firenze, Dipartimento di Fisica e Astronomia, CSDC and INFN, via G. Sansone 1, 50019 Sesto Fiorentino, Italia}
\author{Duccio Fanelli} \affiliation{Universit\`{a} degli Studi di Firenze, Dipartimento di Fisica e Astronomia, CSDC and INFN, via G. Sansone 1, 50019 Sesto Fiorentino, Italia}

\begin{abstract}
Multiple sinks competition is investigated for a walker diffusing on directed complex networks. The asymmetry of the imposed spatial support makes the system non transitive.  As a consequence,  it is always possible to identify a suitable location for the second absorbing sink that screens at most the flux of agents directed against the first trap, whose position has been preliminarily assigned. The degree of mutual competition between pairs of nodes is analytically quantified through apt indicators that build on the topological characteristics of the hosting graph. Moreover, the positioning of the second trap can be chosen so as to minimize, at the same time the probability of being in turn shaded by a thirdly added trap.  Supervised placing of absorbing traps on a asymmetric disordered and complex graph is hence possible, as follows a robust optimization protocol. This latter is here discussed and successfully tested against synthetic data. 
\end{abstract}


\maketitle

\section{Introduction}

Probing the stochastic diffusion on a complex directed network defines a topic of paramount importance and inter-disciplinary breath \cite{barrat, latora}. Depending on the specific realm of 
investigation, and the associated degree of abstraction, one can often identify putative microscopic entities which execute erratic motion inside the region of  pertinence. In many cases, the 
hosting spatial support results in a intricate skeleton of interlinked pathways, which can be effectively mimicked in terms of heterogeneous graphs.  The crowded world of cells \cite{ridgway,kim,luby} is for instance shaped by microtubules, mutually crossing routes that define a nested labyrinth of trails. On a different scale, human mobility flows on veritable networks with asymmetric edges between nodes, as  often roads can be trodden in one direction only. Information flows on Internet, the cyberspace being {\it de facto} schematized as a network with asymmetric routing of the links  \cite{dusi}. In robotics, swarm of  automa \cite{rubenstein} interacts via local and long-ranged exchanges so as to coordinate their respective action.  Swarming behaviors mediated by the diffusive sharing of resources, are also encountered in biological studies of insects \cite{kelley}, ants \cite{goss} and other fields in nature \cite{bonabeau, parisi}. 

Individual constituents, be they molecules, animals or bits of information, stochastically diffusing on the embedding graph should often head to specific targets, ubicated on selected nodes.  
A promotor protein searching for its binding site on DNA \cite{bauer, benichou, mirny}, a web surfer crawling on a chain of hyperlinked pages to reach a given topic of interest \cite{di_patti},  a molecule hunting for the deputed reaction site in topologically tortuous nano-reactors  \cite{y_lu, welsh} or porous media  \cite{benichou_1},  exciton and electron hole recombination or trapping  relevant to photonics and solar-energy science \cite{kopelman}, these are all examples that testify on the widespread significance of devising optimized searching schemes for a stochastic walker on complex geometries assimilated to networks 
\cite{kleinberg, lee, komidis, li_1, li_2, lin, oliveira, oshanin}.

Even more importantly, multiple target sites might coexist and mutually interfere with each other, by screening the flux of incoming ligand particles. Given these premises, 
it is often decisive to elaborate on viable strategies that could yield the most advantageous positioning of a set of target loci (in terms of their associated capturing ability), given a preexisting population of homologous destination sites.  The implications of this fundamental  question are twofold. On the one side, smart positioning of absorbing traps, as target nodes are commonly referred to, could  translate in efficient man made solutions to a large gallery of technological problems. On the other, it might provide a clue to adaptive plasticity in natural phenomena, as shaped by evolution.

To address this topic we shall consider the stochastic dynamics of a walker bound to explore a directed graph, modified with the inclusion of absorbing traps. As we shall argue, the intransitivity of the examined process is a key 
ingredient to the forthcoming analysis.  For pedagogical reasons, we will specialize on a simplified setting where just two traps are considered, although the analysis will extend straightforwardly to graph endowed with an arbitrary number of absorbing sinks.  Assume the first trap to be set a priori on a specific node of the hosting network: is it possible to position the second trap so as to obscure as much as possible the first, and so limiting its capacity to absorb diffusing agents? At the same time, can one minimize the risk that the newly added trap gets in turn weaken by the successive insertion of further absorbing sinks? Inquiring on the aforementioned items implies addressing an optimization problem, that we shall here solve analytically. As we shall argue, the sought optimum depends on the topological characteristics of the scrutinized network, the relevant mathematical quantities depending on the eigenfunctions of the associated discrete Laplacian. The theory will be discussed with reference to a specific family of graphs, which displays the small world effect \cite{WS}. While the optimization problem is trivial and largely uninteresting on directed regular lattices of discretionary connectivity, it is definitely relevant for system hosted on a disordered graph, with long range jumps assigned with a prescribed probability of relocation. Surprisingly, a relatively modest density of long range jumps suffices to yield a meaningful solution to the optimal problem. 

To clarify the potential interest of our conclusions, imagine two competitors that are willing to advertise their own products by flagging them on a node of  a complex asymmetric network, 
e.g. a page on the internet. The first makes her choice and promotes the activity on a specific site, which configures therefore as an absorbing trap for agents (clients) 
unawarely surfing around. Following our recipe, the second investor can place the second trap on a designated node which (i) limits the number of visitors that can reach the site flagged by the opponent and (ii) secure a strategic positioning to reduce the risk of being shaded by other competitors that might join the venture. On a different level, strategies for optimal integration of multiple reactive sites might have been at play, from biology to chemistry, to shape the world the way 
we know it. 

As already mentioned, the optimization scheme to which we alluded above exploits a fundamental property which ultimately stems from having assumed an asymmetric, hence directed, spatial support. In 
the context of game theory this property is termed intransitivity \cite{gardner}. Non-transitive games produces at least one loop of preferences: if strategy A is to be preferred over strategy B, and strategy B 
outperforms strategy C, then strategy A is not necessarily preferred over strategy C. This is for instance the case for the classical rock, paper, scissor game, which is deliberately constructed 
to yield a three steps loop. A more subtle implementation of non transitive game is provided by the so called Penney's game \cite{penney}, a head and tail sequence generating game. 
The first player bets on a binary sequence of assigned length, and discloses it to the second player, who selects in turn another sequence of identical length.  A string is produced by successive tossing of a fair coin, and the player whose sequence appears first, as consecutive readings of the toss outcomes, wins. Provided sequences of at least length three are used, and because of the emerging intransitivity, the second player statistically wins over the starting player: for any given sequence of length three (or longer), another sequence can be always found  that has higher probability of occurring first. Mathematically, the Penney game can be reformulated as a problem of stochastic diffusion on a directed network, whose nodes are the different sequences of fixed length which can be assembled with a binary alphabet. Remarkably, the non-transitivity relates to the asymmetry of the underlying graph, but the two concepts are 
to some extent different, as we shall argue in the following.  As a matter of fact, the analytical treatment here proposed will materialize in a macroscopic indicator to quantify the global  intransitivity of the examined asymmetric graph, enabling one to establish a priori if one contendent can outperform the other or, equivalently if optimal strategies can be played. 

The paper is organized as follows. In the next Section, we will consider the problem of traps competitions for a walker diffusing on a regular asymmetric lattice. Operating the continuum limit 
the problem is mapped into a standard Fokker-Planck equation, in one dimension, subject to two absorbing boundary conditions. The discussion has mainly a pedagogical interest as 
it sets the stage for the forthcoming generalization. As we shall clarify in the following, the optimal problem returns, in this case, a trivial solution. Then, we will turn to 
analytically investigate the competition between fully absorbing traps, positioned on a generic graph, as specified by its adjacency matrix. We will in particular derive closed analytical expressions for the asymptotic density absorbed on each of the mutually competing traps,  and so immediately assess their respective victory rate,  in the spirit of  non-transititive 
games. This information will form the basis for defining the optimization strategy mentioned above.  The proposed approach will be exemplified for a class of graphs of (directed) 
Watts-Strogatz type with a variable long-ranged relocation probability.  
Finally, we shall sum up and conclude. Technical details are illustrated in Materials and Methods.

\section{Results} 

\subsection{Competition between traps on asymmetric regular lattices}
\label{Sect2}

Assume a walker to hop randomly on a one-dimensional directed regular lattice made of $N$ nodes and subject to periodic boundary conditions. Each node is solely connected to its adjacent nearest neighbors. Denote by $a$ (resp. $b$) the probability of jumping towards the right (resp. left). Here, $a$ and $b$ are positive real numbers chosen to match the condition $a+b<1$.
The stochastic $N \times N$ matrix $\bm M$ which controls the diffusive process is therefore circulant, with entries specified by $M_{i,i-1}=b$, $M_{i,i}=1-a-b$, $M_{i,i+1}=a$ in such a way that $\sum_i M_{i,j}=1$,  $\forall j$. Notice that in this preliminary example the edges among connected nodes are symmetric. The asymmetry that makes the graph directed comes from the probability which controls microscopic jumps. In the following Section we shall turn to consider graphs that are topologically asymmetric, namely graphs that present an heterogeneous distribution of links. In all cases, for the sake of simplicity, we shall refer to direct or, equivalently, asymmetric networks.

In the continuum limit, assuming that nodes are densely distributed on the circle, the probability of seeing the walker in a specific spatial location (identified by the continuum variable $x$) si governed by:

\begin{equation}
  \partial_t p(x,t)=-v\partial_x p(x,t)+D\partial^2_x p(x,t)
  \label{FP}
 \end{equation}
where $v=b-a$ and $D=(a+b)/2$ respectively denotes the drift velocity and the diffusion constant\footnote{In performing the continuum limit we are implicitly setting both space and time elementary intervals to unit.}. Working in this context, and given any fixed pair of nodes, $i$ and $j$, we aim at evaluating their relative scores in terms of visits of independent and mutually transparent random walkers. More specifically, we imagine $i$ and $j$ to act as fully absorbing traps. Starting from a uniform distribution (nodes are equally populated at time $t=0$), we wish to estimate the number of paths that take a walker to $i$ (without hitting $j$) and viceversa. This analysis translates into a scalar indicator $V_{ij}$, positive and smaller than one, if properly normalized, that weights the probability of $i$ to {\it win over} $j$. Conversely,  $V_{ji}$ will measure the probability of $j$ to prevail over $i$. Clearly,  $V_{ij}+V_{ji}=1$, as it follows from the obvious conservation of the total probability. In the end, a  $N \times N$ matrix $\bm V$ can be obtained that quantifies the probability 
of every node to win against any other selected competitor site. The diagonal elements of $\bm V$ are arbitrarily set to zero.

\begin{figure}[t]
\begin{center}
\includegraphics[scale=0.5]{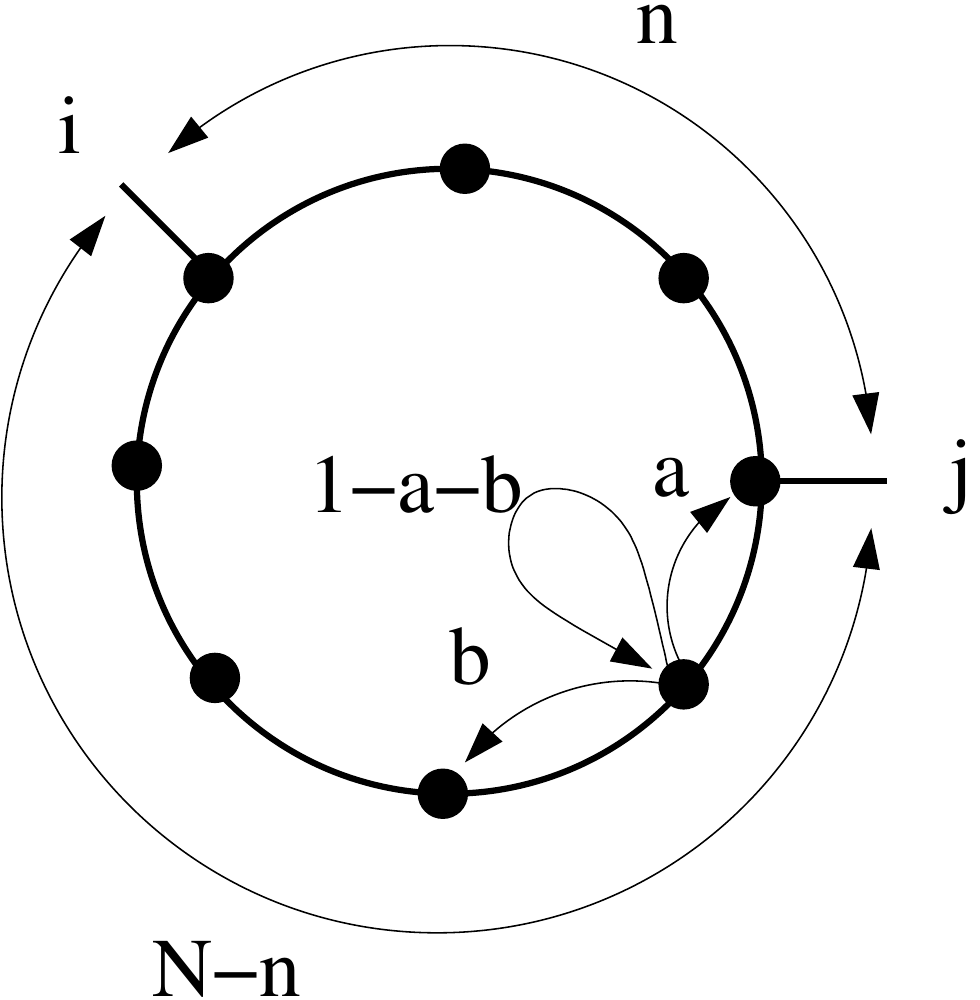} 
\end{center}
\caption{A schematic illustration of the diffusion process on a asymmetric lattice, subject to periodic boundary conditions is provided. The two competing traps are located on nodes $i$ and $j$, respectively. The problem can be equivalently reformulated, by studying the flux of particles inside two adjacent intervals, composed by $n$ and $N-n$ nodes and constrained to match absorbing boundary conditions at the edges. \label{figure1}}
\end{figure}

To determine explicitly the matrix $\bm V$ we first position two fully absorbing traps in respectively node $i$ and $j$. Then, we divide the circle (regular lattice under periodic boundary conditions) into two distinct domains, as schematically depicted in Figure  \ref{figure1}. One domain is constituted by the $n$ nodes encountered when circulating from $i$ to $j$, clockwise. The ensemble made by the complementary $N-n$ nodes (from $j$ to $i$, clockwise) defines the second set. To quantify the asymptotic density of walkers that fall on each trap, we need to solve the Fokker-Planck equation
(\ref{FP}), inside both domains and subject to absorbing boundary conditions at their respective edges. More specifically, we will consider the general solution of the Fokker-Planck equation defined on a one dimensional segment $[0,L]$, with $p(0,t)=p(L,t)=0$. Further, we shall begin by assuming as initial condition a Dirac delta centered in $x_0=\alpha L$, with $0<\alpha<L$. Following \cite{soluzFP} (see Materials and Methods), the sought solution reads:

 
 \begin{equation}
  p(x,t) = \frac{\Gamma}{2L}e^{\frac{v}{2D}(x-\alpha L)-\frac{v^2}{4D}t}  
  \label{p(x,t)}
 \end{equation}
 
 where
 
\begin{equation}
\Gamma \equiv \biggl[\theta_3\biggl(\frac{\pi}{2}(\alpha-\frac{x}{L}),z(t)\biggr)-\theta_3\biggl(\frac{\pi}{2}(\alpha+\frac{x}{L}),z(t)\biggr)\biggr]
 \end{equation}
 
 and: 
 
 \begin{equation}
  \theta_3(r,q)\equiv1+2\sum_{k=1}^{\infty}\cos(2rk)q^{k^2} \qquad z(t)\equiv e^{-\frac{\pi^2Dt}{L^2}}.
 \end{equation}
 
 Here $\theta_3(\cdot,\cdot)$ stands for the Jacobi theta function.
We are now in a position to evaluate the probability current $J(x,t)=-D\partial_x p(x,t)+vp(x,t)$, 
flowing to the boundaries, namely   $J(L,t)$ e $J(0,t)$. As outlined in Materials and Methods, one eventually gets:

 \begin{equation}
  J_{\rightarrow}(t)\equiv J(L,t)=\frac{D\pi}{2L^2}e^{\frac{v}{2D}L(1-\alpha)-\frac{v^2}{4D}t}\theta_3'\biggl(\frac{\pi}{2}(\alpha+1),z(t)\biggr)
  \label{JL}
 \end{equation}
 \begin{equation}
  J_{\leftarrow}(t)\equiv -J(0,t)=-\frac{D\pi}{2L^2}e^{-\frac{v}{2D}\alpha L-\frac{v^2}{4D}t}\theta_3'\biggl(\frac{\pi}{2}\alpha,z(t)\biggr)
  \label{J0}
 \end{equation}
 where $\theta_3'(\tilde{r},q)=\partial_{r}\theta_3(r,q)|_{r=\tilde{r}}$, and where we have introduced the positive quantity  $J_{\rightarrow}(t)$  (resp. $J_{\leftarrow}(t)$) to denote the 
current flowing from the right (resp. left) boundary. The probability that particles get absorbed to either right ( $\mathcal{J}_{\rightarrow}$) or left ( $ \mathcal{J}_{\leftarrow}$) boundary is obtained by respectively integrating $J_{\rightarrow}(t)$ and $J_{\leftarrow}(t)$ to yield:

\begin{equation}
  \mathcal{J}_{\rightarrow}(L)=\int_0^{\infty}J_{\rightarrow}(t)dt=\frac{1-e^{-\frac{\alpha v L}{D}}}{1-e^{-\frac{v L}{D}}}
 \end{equation}
 \begin{equation}
  \mathcal{J}_{\leftarrow}(L)=\int_0^{\infty}J_{\leftarrow}(t)dt=\frac{e^{-\frac{\alpha v L}{D}}-e^{-\frac{v L}{D}}}{1-e^{-\frac{v L}{D}}}
 \end{equation}
 
To estimate the relative performance of the two traps $i$ and $j$, we have to generalize the analysis to the case of an initial uniform distribution of the walkers on the lattice. For this reason, we shall integrate the above expressions over the allowed interval in $\alpha$ to yield:

 \begin{equation}
  \mathcal{V}_{\rightarrow}=\int_0^1 \mathcal{J}_{\rightarrow}d\alpha=\frac{1}{1-e^{-\frac{v L}{D}}}-\frac{D}{vL}
 \end{equation}
 \begin{equation}
  \mathcal{V}_{\leftarrow}=1-\mathcal{V}_{\rightarrow}=1-\frac{1}{1-e^{-\frac{v L}{D}}}+\frac{D}{vL}.
 \end{equation}

These preliminary relations will be used to assess the relative performance of the two traps $i$ and $j$. We recall that the circular lattice that defines the spatial background of the model has been split into two distinct domains: the first formed by the $n$ nodes, visited when going from $i$ to $j$, clockwise.  The second domain is constituted by the remaining $N-n$ nodes, encountered when circulating the ring clockwise from $j$ to $i$. With reference to the former, $\mathcal{V}_{\rightarrow}$ stands for the flux of particles that eventually hits $j$, while $\mathcal{V}_{\leftarrow}$ refers to the particles that are eventually 
attracted towards $i$. For the other domain, the situation is clearly specular. Hence, the probability $V_{ij}$ that a diffusing agent is eventually attracted to node $i$ instead of node $j$ (i.e. without passing from node $j$) is the sum of two terms: $\mathcal{V}_{\leftarrow}$ calculated for $L=n$ and  $\mathcal{V}_{\rightarrow}$ with $L=N-n$. The first term should be weighted by a factor $n/N$ to reflect the average over the initial uniform distribution, while the second needs to be multiplied by a factor $(N-n)/N$. In formulae:

\begin{equation}
 \begin{split}
 V_{ij}  &=\frac{n}{N}\mathcal{V}_{\leftarrow}(n)+\frac{N-n}{N}\mathcal{V}_{\rightarrow}(N-n)=\\
  &=\frac{n}{N}-\frac{n}{N}\frac{1}{1-e^{-\frac{v}{D}n}}+\frac{N-n}{N}\frac{1}{1-e^{-\frac{v}{D}(N-n)}}.
  \end{split}
  \label{Si}
 \end{equation}
 
 With analogous consideration one gets:
 
 \begin{equation}
 \begin{split}
 V_{ji} &=\frac{n}{N}V_{\rightarrow}(n)+\frac{N-n}{N}V_{\leftarrow}(N-n)=\\
  &=\frac{N-n}{N}+\frac{n}{N}\frac{1}{1-e^{-\frac{v}{D}n}}-\frac{N-n}{N}\frac{1}{1-e^{-\frac{v}{D}(N-n)}}.
  \end{split}
  \label{Sj}
 \end{equation}

For $v/D \simeq 0$ diffusion prevails over drift: on average, half of the particles are expected to fall on trap $i$ and the remaining ones to get absorbed by trap $j$.  This is in turn the limit of a symmetric adjacency matrix for the investigated stochastic dynamics, that yields $V_{ij}=V_{ji}=1/2$. At variance, for large values of the ratio $v/D$, the exponential functions in equations    
(\ref{Si}) and (\ref{Sj}) can be neglected and one eventually obtains  $V_{ij} \simeq 1-n/N$, and, obviously $V_{ji} \simeq n/N$. 

The intransitivity ultimately stems from the underlying network asymmetry, here implemented through unbalanced jumping rates. Although related, the concepts of asymmetry and intransitivity are however subtly different. 
When the adjacency matrix is asymmetric, it is not a priori guaranteed that, for any selected node $i$, at least another node $j$ exists that wins over $i$. Stated differently, for a generic random walker hopping on a directed graph, the entries of a given column(s) of matrix $\bm V$ can be in principle smaller than $1/2$. In the following, we are interested in identifying a specific subclass of asymmetric networks, that we shall term globally intransitive. For these networks, any trap $i$ can be always (statistically) outperformed, in terms of its ability to absorb, by at least another trap, positioned on a given node $j$. To formally classify the asymmetric network according to this scheme, we introduce an {\it index of global intransitivity}, $\eta$, calculated via the following procedure. We select the maximum from each column of $\bm V$, and then identify the global minimum among collected values. This latter quantity is then shifted by $-1/2$, to yield the index $\eta$, which is therefore bound to the interval $[-1/2,1/2]$. If $\eta$ is positive the system is globally intransitive, according to the definition evoked above. Classical measures of networks intransitivity rely on triad census. The transitivity coefficient of a network, often termed {\it clustering coefficient}, is the ratio of the number of loops of length three and the number of paths of length two \cite{latora}. In other words, the clustering coefficient quantifies the frequency of loops of length three in the network. The parameter $\eta$ returns instead a more general estimate of the intransitivity degree, as it does require assuming a priori a specific size of underlying loops. 

For the case under scrutiny of a regular asymmetric lattice, $\eta \rightarrow 0$, when the drift is virtually silenced ($v/D \rightarrow 0$) and the process approaches the symmetric limit. Conversely, for $v/D  \ne 0$,  $\eta > 0$ and it approaches the limiting value  $\eta \rightarrow 1/2$ for  $v/D  \rightarrow \infty$. The process of asymmetric particles' hopping on a regular lattice with short ranged connections is therefore globally intransitive for any $v/D  \ne 0$.

Since the process is always globally intransitive, a secondly added trap can always be found that wins over the first,  by attracting more walkers. Is it however possible to
position the second competing trap ($j$) on a node that reduces as much as possible the risk of being obscured by yet another trap, the third of the sequence, while still performing better than the first ($i$)? To answer this question, we take advantage of the composite information stocked inside matrix $\bm V$. We introduce in fact an additional indicator, called $\sigma_{ij}$ and defined as follows:

\begin{equation}
\sigma_{ij} = \frac{\sum_{k \ne i} V _{jk}}{N-2}
 \label{sigma}
 \end{equation}

The larger the value of $\sigma_{ij}$ the less the average screening on trap $j$ (selected after trap $i$), as exerted by an hypothetical third trap, installed in one of the remaining $N-2$ nodes of the network. Notice that, by definition, $\sigma_ {ij}$ stays in the interval $[0,1]$.  

\begin{figure}[t]
\begin{center}
\includegraphics[scale=0.35]{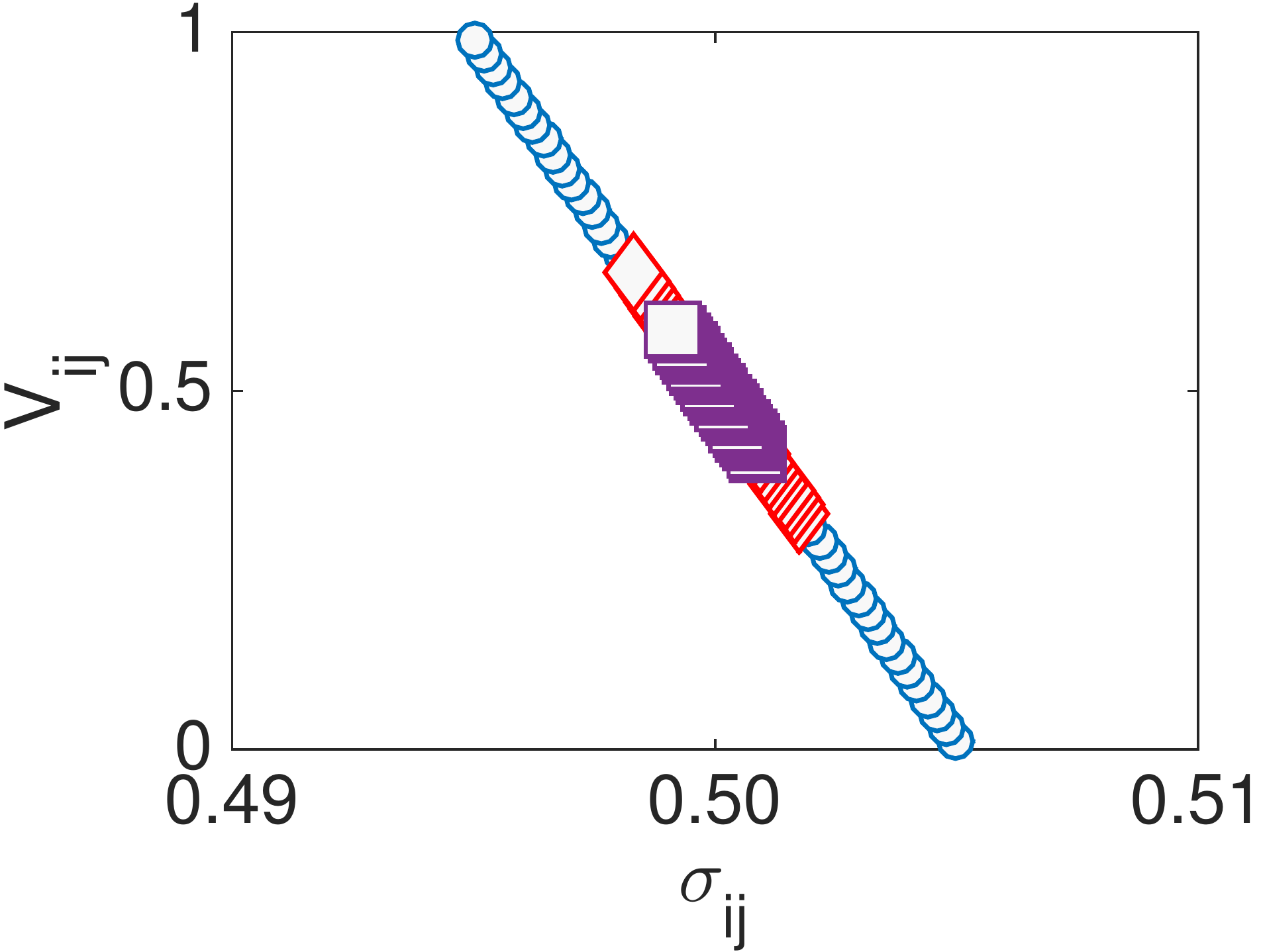} 
\end{center}
\caption{(Color online) $V_{ij}$ vs. $\sigma_ {ij}$ as obtained for regular asymmetric lattices of the type schematized in Figure  \ref{figure1}. Violet squares refer to $v/D=0.01$, red diamonds to $v/D=0.02$ and blue circles to $v/D=2$. As  $v/D$ gets larger the distribution of points stretches vertically. To favor visualization we have chosen to plot a subset of the full data points list, for each choice of $v/D$. Notice the horizontal scale: the points align almost vertically.  
 \label{figure2}}
\end{figure}

Building on the above we are now in a position to define a supervised strategy for optimizing the selection of trap $j$, given the pre-defined location of trap $i$ and for a stochastic diffusion process, 
taking place on a globally intransitive graph. For fixed $i$, the key idea is to select $j$ in such a way that it both maximizes $V_{ij}$, the factor that quantifies direct competition versus $i$,  and  $\sigma_ {ij}$, a measure of competition against the residual bulk. In Figure  \ref{figure2}   $V_{ij}$ is plotted versus $\sigma_ {ij}$ for regular one-dimensional asymmetric lattices, characterized by increasing values of $v/D$ amount. Different symbols refer to different 
choices of $v/D$. In carrying out the analysis we considered all possible combination of $i$ and $j$. As it can be clearly appreciated by visual inspection, the data align on a almost vertical line, the value of $\sigma_ {ij}$ being, for practical purposes, constant. The larger the value of $v/D$, the wider the vertical band. Optimizing the selection of node $j$, given $i$, proves therefore a trivial exercise, when the system is made to diffuse on a regular directed lattice: the best choice is to select the node $j$ which maximizes the $V_{ij}$ score, irrespectively of the corresponding    
$\sigma_ {ij}$. The newly introduced trap will be manifestly fragile, as concerns the successive intrusion of additional traps. As we shall see in the following section,  complex asymmetric  
networks, that accomodate for directed long-range jumps to distant sites, yield however a definitely richer scenario and, consequently, more intriguing optimization protocols. 

Before concluding this Section, we remark that, in the limit  $v/D \rightarrow \infty$, the matrix element  $V_{ij} \simeq 1-n/N$. Hence,  $\sigma_{ij}$ as defined in (\ref{sigma}) can be calculated analytically to return  $\sigma_{ij}= (N-3)/(N-2)/2+n/(N-2)/N$, where $n$ refers to the number of sites entrapped  in between node $i$ and $j$, see Figure \ref{figure1}. This latter estimate accurately explains the peculiar distributions as seen in Figure \ref{figure2}.

\subsection{Traps on a directed disordered network and the optimization problem}
\label{Sect3}

We here aim at extending the above analysis to the case of a walker that is randomly diffusing on a generic directed network. The entries of the adjacency matrix
are one, if two nodes are mutually connected, or zero otherwise. Hence, at variance with the case considered above, the asymmetry of the support is topological, namely 
related to the distribution of assigned edges, while the probability of individual jumps is constant and set to one, without loss of generality. In the following, we shall denote with $\bm M$ the stochastic matrix obtained by
dividing the columns of the adjacency matrix by the associated nodes connectivity.

As discussed above we are interested in resolving the degree of mutual interference between two distinct traps, respectively located in $i$ and $j$. Walkers can reach the absorbing traps, but cannot escape from them.  To accommodate for this effect we replace the $i$-th  and $j$-th columns of matrix $\bm M$ with zeros, except for the diagonal elements which are instead set to one. The obtained matrix is hereafter referred to as to $\bm M^{[i,j]}$. The master equation that governs the evolution of $p_k$, the probability of detecting a particle on node $k$, reads \footnote{Again we implicitly assume discrete time updates with $\Delta t=1$.}:

   \begin{equation}
   \dot{p}_k(t)=\sum_l M^{[i,j]}_{kl} p_l(t)-\sum_l M^{[i,j]}_{lk} p_k(t) = \sum_l L^{[i,j]}_{kl} p_l(t)
   \label{ME}
   \end{equation}
 
where $L^{[i,j]}_{kl}= M^{[i,j]}_{kl}-\delta_{kl}$ is the Laplacian operator in presence of absorbing traps. The last equality in (\ref{ME}) follows immediately from the normalization condition  $\sum_l M^{[i,j]}_{lk}=1$. To solve the linear problem (\ref{ME}) we need to develop the time dependent probability on a proper basis, which diagonalizes the Laplacian operator. The associated eigenvalue problem takes the form $\sum_{l} L^{[i,j]}_{kl} \psi_l^{(\alpha)} =  \lambda^{(\alpha)}  \psi_k^{(\alpha)}$ where  $\lambda^{(\alpha)}$ and $\bm \psi^{(\alpha)}$ define, respectively, the eigenvalue and its associated, $N$-dimensional, eigenvector. It can be proven that two eigenvalues  of the discrete Laplacian operator exist which are identically equal to zero and that reflect the imposed absorbing traps.  Importantly, and because of the asymmetry of the network, all remaining eigenvalues are complex, and bear a negative real part. The eigenvectors that correspond to null eigenvalues have a rather simple structure: all their components are zero, except for the entry identified by the consonant trap index. This latter component is equal to one. The eigenvectors associated to the Laplacian operators are linearly independent, but they do not constitute an orthonormal basis, as it instead happens when the underlying graph structure is supposed to be symmetric. To solve  the linear equation  (\ref{ME}) we shall preliminarily define an appropriate orthonormal basis $\{\bm u^{(\beta)}\}$, expressed in terms of the original eigenvectors  $\bm \psi^{(\alpha)}$ of the Laplacian operator.  As we shall see, the request of dealing with an orthogonal basis is fundamental to carry out the forthcoming derivation. Mathematically, one can always find a linear transformation  such that $u_k^{(\beta)}=\sum_{\alpha}C_{\alpha\beta}\psi_k^{(\alpha)}$, where $\bm C$ is the $N \times N$ matrix that specifies the change of basis. We can hence set: 

\begin{equation}
 p_k(t)=\sum_{\beta}\hat{p}_{\beta}(t)u_k^{(\beta)}
 \label{espans2}
\end{equation}
where $\hat p_{\beta}$ represents the coefficients of the expansion on the introduced orthonormal basis. Inserting the above ansatz into equation (\ref{ME})  and carrying out the calculation, that we discuss in some details in the  Materials and Methods Section, one eventually obtains the following explicit solution: 
 \begin{equation}
 p_k(t)=\sum_l p_l(0)\sum_{\beta}(u_l^{(\beta)})^* \sum_{\alpha}C_{\alpha\beta}\psi_k^{(\alpha)}e^{\lambda^{(\alpha)}t}.
 \label{sol_rho}
\end{equation}
The asymptotic solution  $p_k^{\infty}$ that is relevant for our purposes can be readily obtained, by performing the limit for $t \rightarrow \infty$  in  (\ref{sol_rho}), and so yielding:
 
  \begin{equation}
  p_k^{\infty}=\sum_l p_l(0)\sum_{\beta}(u_l^{(\beta)})^*\biggl[C_{i\beta}\psi_k^{(i)}+C_{j\beta}\psi_k^{(j)}\biggr]
  \label{asy}
 \end{equation}
 where use has been made of the fact that two eigenvalues (those associated to the traps) are zero and all the other have negative real parts (their contributions, stored in the exponential, fade away in the large time limit). In the above equation $(\cdot)^*$ stands for the complex conjugate. Recall now that $\psi_k^{(i)}=\delta_{ik}$ e $\psi_k^{(j)}=\delta_{jk}$, which allow to further simplify equation  (\ref{asy}) as:

  \begin{equation}
  p_k^{\infty}=\sum_l p_l(0) \sum_{\beta}(u_l^{(\beta)})^*\biggl[C_{i\beta}\delta_{ki}+C_{j\beta}\delta_{kj}\biggr]
  \label{sol_as}
 \end{equation}
 
 Eventually the walker has to land either on site $i$ or $j$, where the absorbing sinks are located. The relative ability of $i$ and $j$ to trap stochastically diffusing entities is quantified through the following elements of matrix $\bm V$:
 
  \begin{equation}
  V_{ij}=\sum_l p_l(0)\sum_{\beta}(u_l^{(\beta)})^*C_{i\beta}
 \end{equation}
  
  and
  
  \begin{equation}
  V_{ji}=\sum_k p_l(0)\sum_{\beta}(u_k^{(\beta)})^*C_{j\beta}
 \end{equation}

with $p_l(0)=1/N$, in the relevant case where the initial condition is assumed to be uniform.  Having determined the elements of the matrix $\bm V$, we are in the position to estimate both the global intransitivity index $\eta$ and the parameter $\sigma_{ij} $ as defined in the preceding Section. As we will demonstrate in the following, the optimization problem discussed above admits a non trivial solution, when the hosting  graph is heterogeneous, and as opposed to the simplified setting where the diffusion occurs on a regular asymmetric lattice \footnote{Incidentally, we note that the above relations provide closed analytical solutions to the family of Penney's games}. 

\begin{figure}[t]
\begin{center}
\includegraphics[scale=0.3]{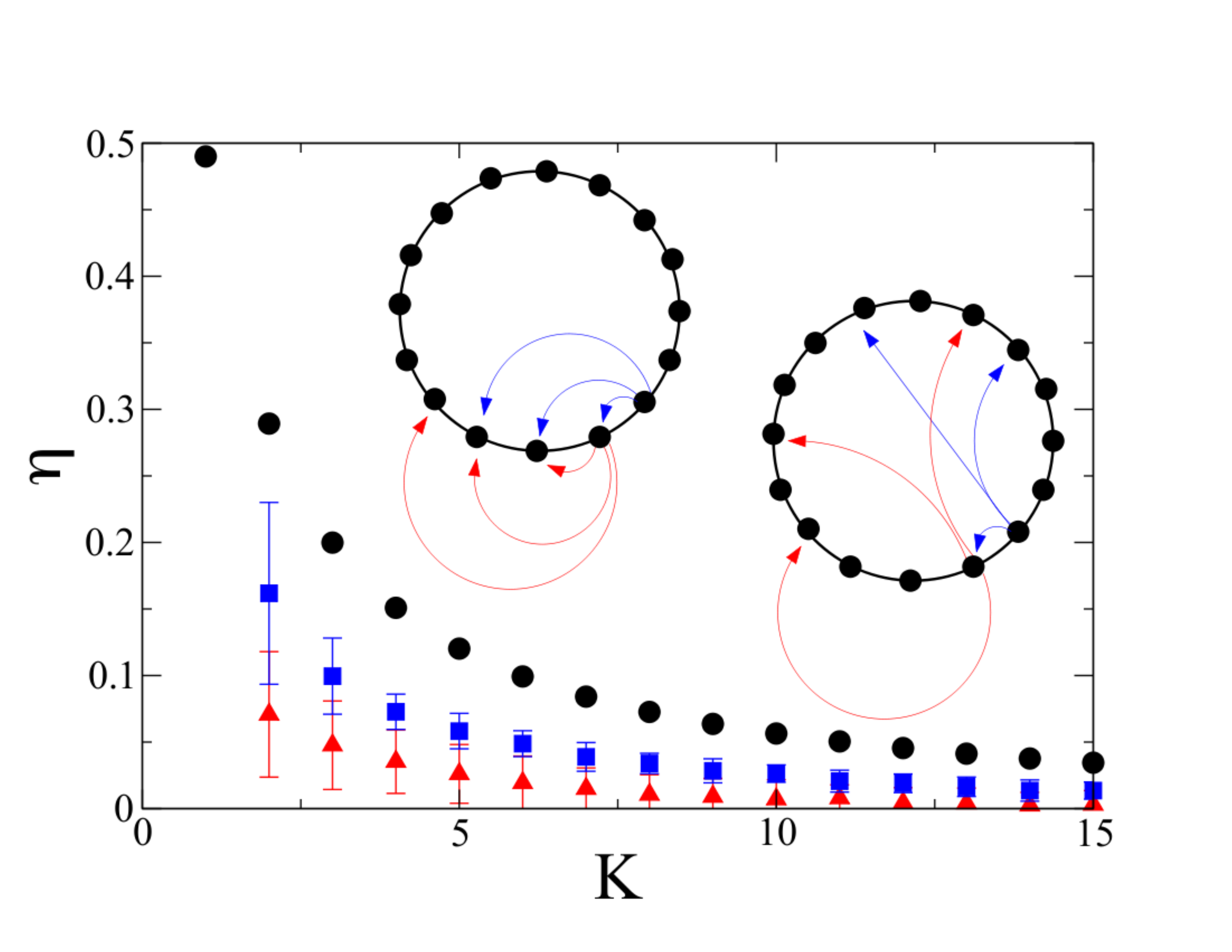} 
\end{center}
\caption{The index of global intransitivity $\eta$ is plotted as a function of the (outgoing) connectivity $K$ for the case of regular $K$-ring lattices (circles) and disordered graphs, with $p=0.01$ (squares) and $p=0.1$ (triangles), respectively. In all cases $N=100$. The data obtained for the disordered graphs have been averaged over $100$ independent realizations. The errors are the recorded standard deviations. The insets provide a pictorial illustration of a $K$-regular lattice (left) and a disordered network (right) \label{figure3}}
\end{figure}

To prove our claim we consider a family of directed graphs generated via a straightforward procedure which is adapted from the Watts-Strogatz recipe.   
Assign the desired number of nodes $N$ and be $K$ their assigned constant degree (connectivity). We then construct a $K$-regular ring lattice, by connecting each node to its $K$ nearest neighbors, on one side only. Then, for every node $i$, we select all its associated edges and rewire them with a given probability $p \in [0, 1]$. Rewiring implies replacing the target node, with one of the other nodes, selected with a uniform probability from the ensemble of possible destination sites. The rewiring is directed and the outgoing connectivity is preserved. In Figure \ref{figure3} we report $\eta$ as a function of the connectivity $K$, for (i) the $K$-ring lattice, and the corresponding disordered graph obtained by imposing a different probability of rewiring, respectively (ii) $p=0.02$ and (iii) $p=0.2$. In all cases the index $\eta$ is positive, hence implying that the system is globally intransitive, a prerequisite condition for the optimization protocol to be applicable. Moreover $\eta$ decreases as $K$ is increased and, more importantly, as $p$ gets larger. The more disordered the graph, the less intransitive the network appears at the global scale, a reasonable result as rewiring amounts to breaking the perfect asymmetry of the initial lattice and so enforcing a macroscopic symmetrization in the topology of the hosting support. 

\begin{figure*}[t]
\begin{center}
\begin{tabular}{cc}
\includegraphics[scale=0.35]{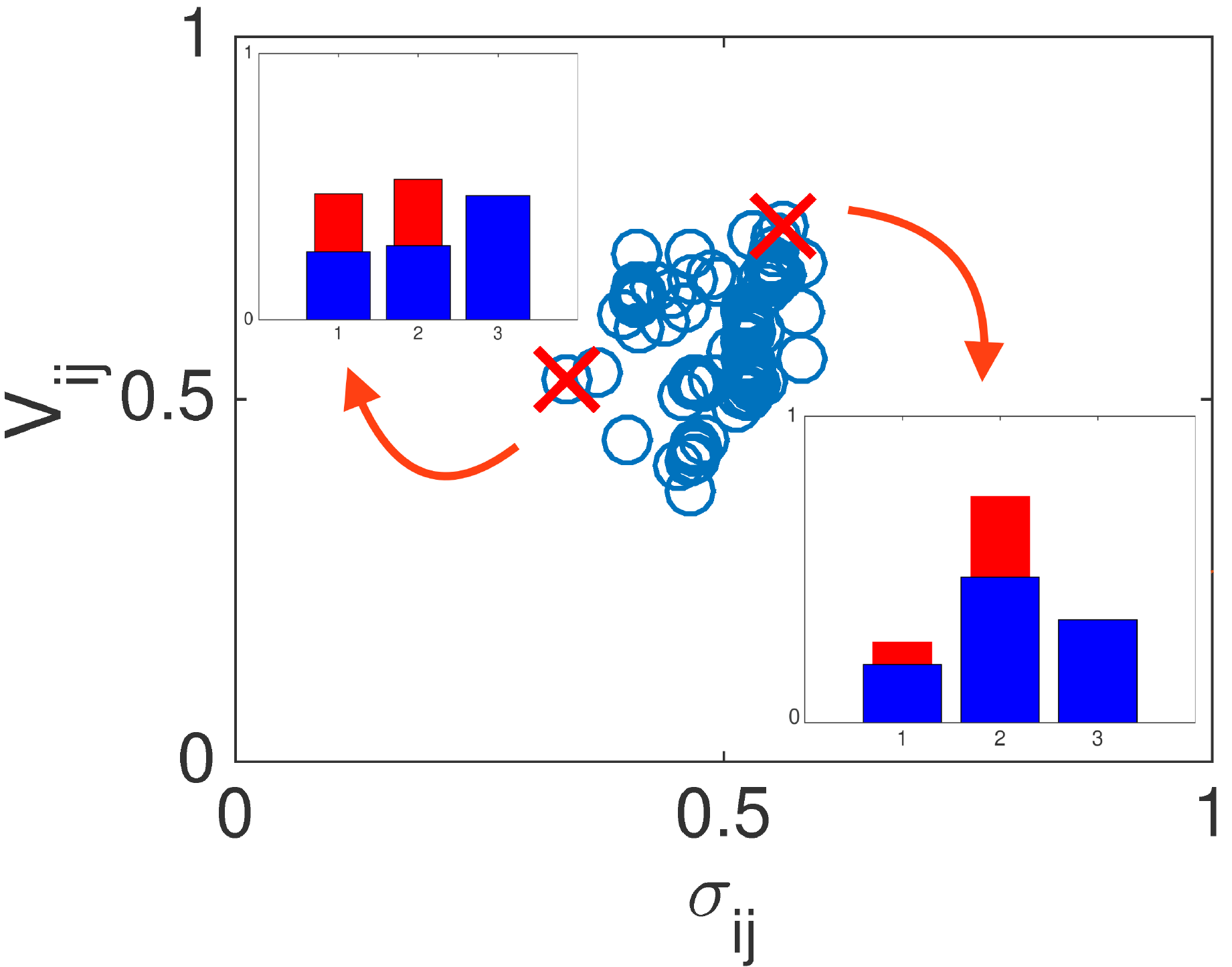} &
\includegraphics[scale=0.35]{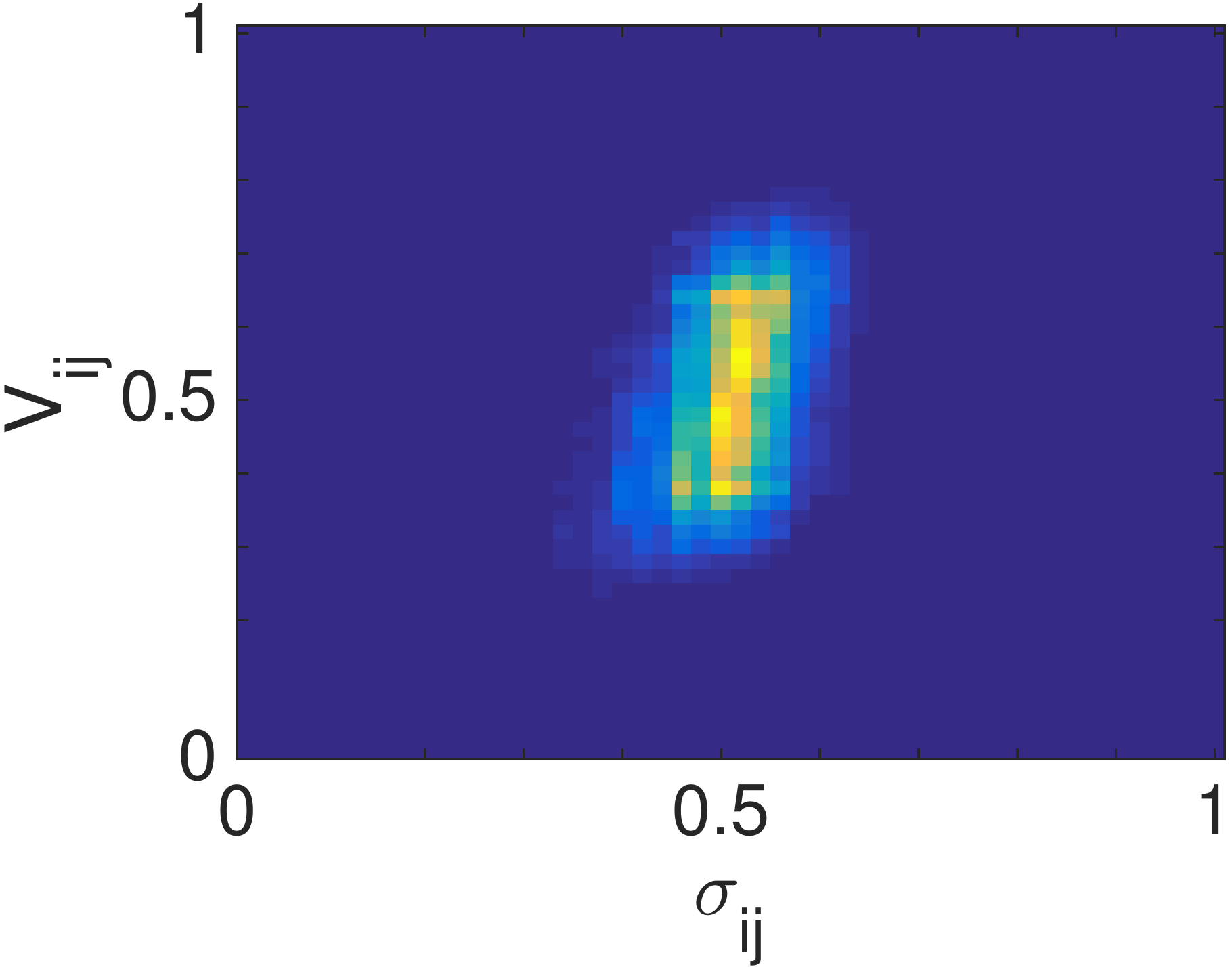}\\
\includegraphics[scale=0.35]{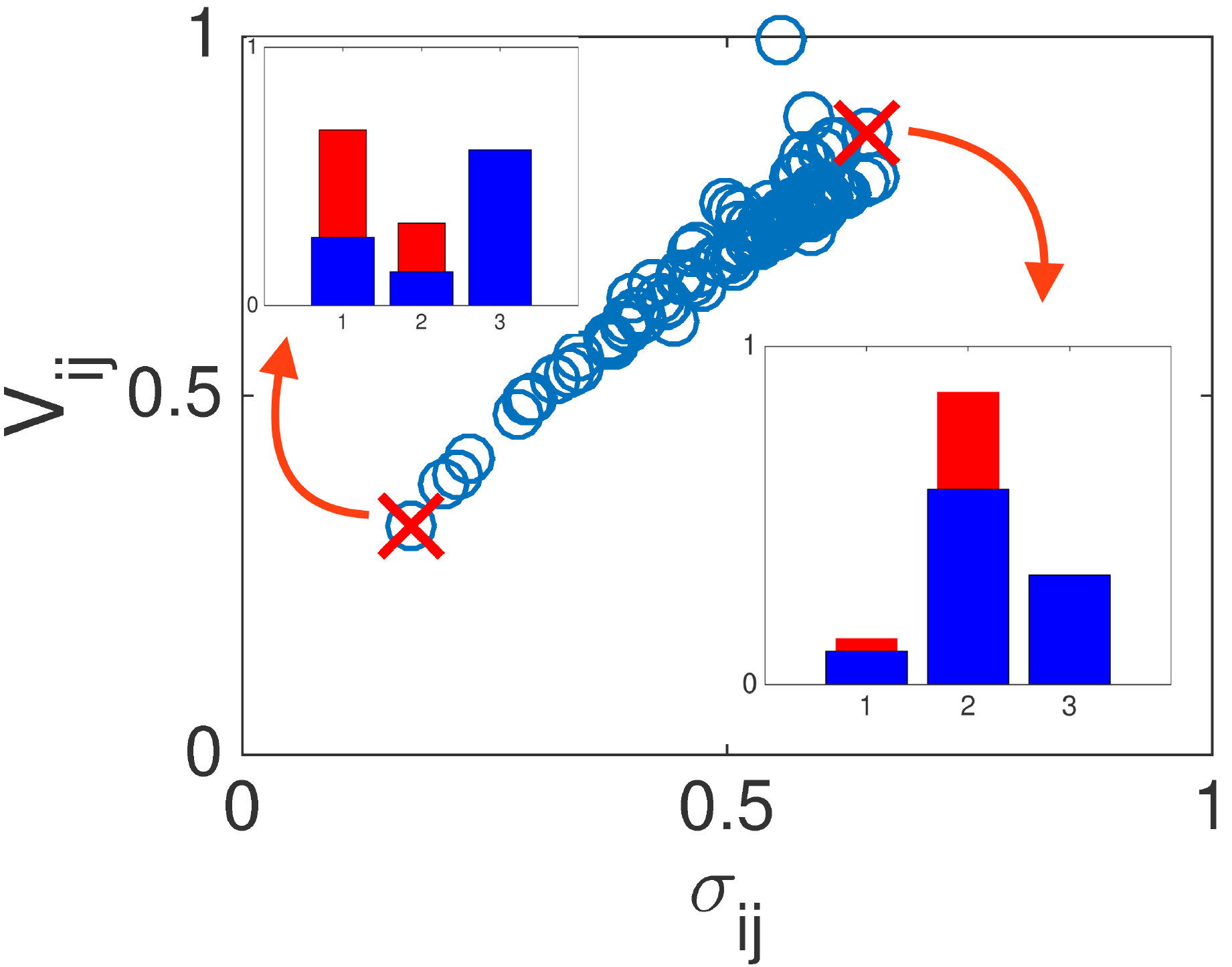}&
\includegraphics[scale=0.35]{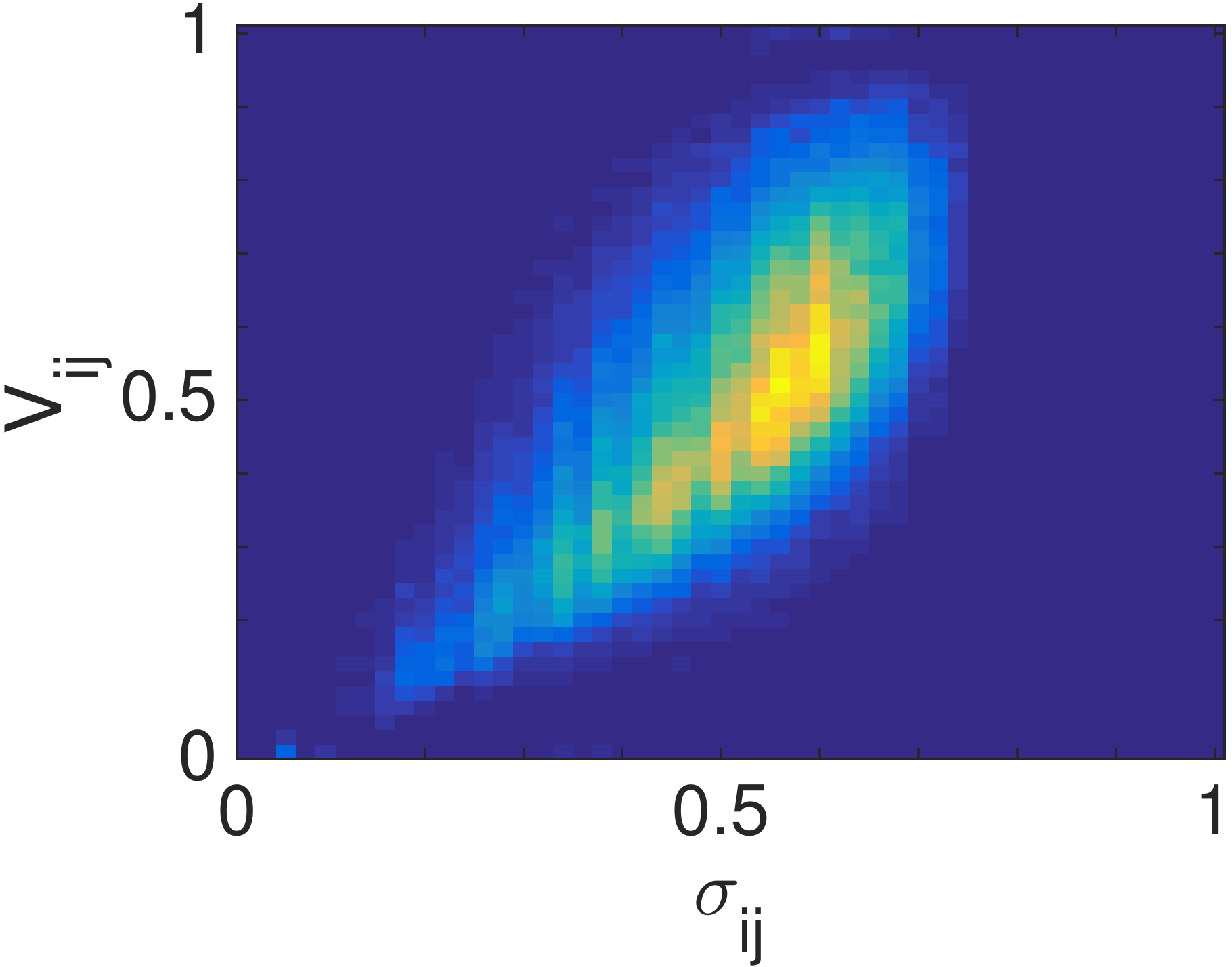}\\
\end{tabular}
\end{center}
\caption{(Color online) Top left panel:  $(V_{ij}$ vs. $\sigma_ {ij})$ for a specific choice of $i$, and running $j$ on the remaining $N-1$ nodes. Here, $p=0.02$. Two different choices of $j$ are evidenced which 
correspond to optimal (top-right) and un-optimal (lower-left) selections, according to the criteria illustrated in the main body of the paper. The histograms report on the performance of the traps $i=1$ and $j=2$, when these are the sole sinks present  (red thin bars) and when they are competing with trap number $3$ (blues tick bars). The data are calculated by averaging over the $N-2$ possible locations for the third trap. The lower histogram photographs the optimal scenario (rightmost red cross). The upper histogram corresponds to the sub-optimal condition (leftmost red cross). Top right panel: density plot for $(V_{ij}$ vs. $\sigma_ {ij})$, 
as obtained for a K-ring with $p=0.02$, upon averaging over $20$ independent realizations  and allowing all possible selection for the first trap $i$. Lower panels: same as for the top panels, with $p=0.2$. Here, N=100.
As it can be clearly appreciated, the distribution of points gets progressively stretched along the diagonal, making the proposed optimization protocol gradually more effective for increasing value of the long-ranged distortion $p$. 
Notice that a modest probability of relocation $p$ suffices to yield an interesting optimization, an intriguing observation which can be ascribed to the peculiarities of small world networks. 
\label{figure4}}
\end{figure*}

We now turn to consider the optimization process, following the approach illustrated in the preceding Section. We recall that the idea is to select the location of the second trap 
$j$, after having fixed the first one $i$, so as to maximize, at the same time, $V_{ij}$, the measure of direct competition versus $i$,  and  $\sigma_ {ij}$, the quantity that controls the degree of competition against the remaining $N-2$ nodes. The results of analysis are reported in Figure \ref{figure4}, for two different choices of the parameter $p$ (top and lower panels, respectively). For a given network realization, we select a generic node $i$, which identifies the location of the first trap. The $N-1$ symbols scattered in the plane  $(\sigma_ {ij}, V_{ij})$ (panels in the left) gauge the performance of the other nodes, imagined as the competitor trap $j$.

Performing the same analysis for the limiting case $p=0$, returns a  distribution that is substantially uninteresting in the perspective of devising a viable optimization strategy, consistently with the analysis carried out in the preceding Section (data not shown). For $p \ne 0$, instead, the distribution of points  gets distorted and progressively elongates along the bisectrix, as clearly testified by visual inspection of Figure \ref{figure4}.
Remarkably, the more the global intransitivity index $\eta$ gets reduced, the more the density points tend to populate the top-right portion of the parameter plane, which incidentally identifies the region of interest for the optimization  method here addressed. To cast it differently, when the graph becomes disordered, while still being asymmetric, it is definitely possible to operate a supervised selection of an absorbing sink $j$, for {\it any} given choices of $i$, that outperforms the latter in terms of ability to attract and, still, minimizes the risk of being in turn buried by successively added traps. The rightmost panels of Figure \ref{figure4} display the density plot obtained upon averaging over $20$ independent realizations of the networks, generated with an assigned rewiring probability $p$. The trend agrees with the general conclusion illustrated above for a specific choice of $i$. It is remarkable that a modest amount of long-range edges suffice to yield a significant optimization scenario. This is made clear in the annexed movie, where the density plots are displayed for increasing values of $p$, in a range of definition that returns a sensible optimization scheme. When $p$ exceeds a given threshold, the networks is too unstructured to allow for a global optimization, even if local strategies can be played in light of the persistent  grade of intransitivity.  

Before ending this Section, we wish to assess the effectiveness of the proposed method. With reference to the choice $p=0.02$, we place the second trap in the optimal position, as identified in the top left panel of Figure \ref{figure4}. We evolved numerically the stochastic dynamics of the system, starting from a uniform initial distribution, and found that trap $j$ captures almost $75 \%$ of the diffusing agents, the remaining ones heading to $i$. But what is going to happen when a third trap is introduced into the competition? Averaging over the $N-2$ possible locations of the third trap, we see that the second trap still has the lion's share with about $50$ $\%$ visits out of the total. Conversely, when the second trap $j$ is assigned to the sub-optimal position, as highlithed in Figure \ref{figure4}, it is solely invested by $15$ $\%$ of the total flux, the remaining quota being directed towards the other two competing sinks. These results are schematically summarized in the insets of the top-left panel of Figure \ref{figure4}. For the case $p=0.2$, a similar scenario holds: the optimal trap $j$  scores  $85$ $\%$, while trap $i$ displays only  $15$ $\%$. Inserting the third trap proves mainly at the detriment of the first sink, the second winning the competition with a final $60$ $\%$ score. If trap $j$ is assigned to its sub-optimal configuration, as depicted in Figure \ref{figure4} (lower-left panel), the final score is as expected very modest $15$ $\%$. The histograms inserted in the left-lower panel of Figure \ref{figure4} summarizes these results.
Finally, we varied $j$, among those nodes that display similar $V_{ij}$ entries, for $i$ fixed.  Reducing $\sigma_{ij}$ is indeed beneficial, as anticipated by our interpretative scheme.


\section{Conclusions}
\label{Sect4}

Complex webs of interactions between individual constituents are everywhere, from the Internet to epidemic spreading, via the molecular processes occurring inside a cell.  Hence, studying the stochastic diffusion of microscopically entities confined on a heterogeneous and disordered spatial support defines a topic of investigation of broad applied and fundamental interest. In many cases, randomly hopping agents are  
chasing for specific target sites, where reactions can supposedly take place. These latter sites configure as absorbing traps. When multiple sinks are simultaneously present on a given network, they can eventually 
interfere with each other,  and mutually screen the flux of incoming ligand agents. Optimization strategies are in principle possible that could result on the most advantageous positioning of an ensemble made of competing traps, so as to minimize their reciprocal hindrance and favor cooperation. On the other hand, it could be strategically advantageous to maximize the mutual competition between traps, so that the lastly added sink 
prevails over previously existing ones. Shedding light onto such an issue can contribute to explain a large plethora of natural phenomena, as shaped by the evolutionary pressure and devise novel efficient man made solutions to specific  technological problems.  

In this paper we have specifically considered the stochastic dynamics of a walker diffusing on a directed static graph, endowed with two absorbing sinks.  Given the network, we imagined the location of the first trap to be assigned a priori. The position of the second trap is designated so as to obscure as much as possible the first, upon estimation of a quantitative indicator that characterizes the degree of pair-interference.  At the same time, an optimal location of the second absorbing sink can be determined that allows to minimize the average screening due to a newly added (third) trap. Analytical formulae are derived which implicitly depend on the topology of the scrutinized network and that enable us to tackle the above optimization process. For walkers diffusing on a regular lattice, and subject to a constant drift,  the optimization protocol is largely ineffective: for any given trap $i$, a competitor sink $j$  can always be found that absorbs a substantial amount of incoming flux of diffusing agents, at the detriment of $i$. The average screening coming from an hypothetic third trap is however relevant and substantially independent on the specific location of the assigned trap $j$.  As opposed to this conclusion, the proposed optimization strategy is definitely significant when the random walker is made to diffuse on a disordered directed graph, with long range relocation edges.  A modest degree of disorder suffices to yield an effective optimization scheme, as we demonstrated in the paper for  specific family of asymmetric complex networks of the Watts Strogatz type. Explicit formulae are derived which materialize in a new class of indicators for the topological characteristic of complex random graphs.
Interestingly, and as a side result, we also propose a global measure for the grade of intransitivity of  a network, which prescinds from identifying closed loops of a given size, as customarily done. 
The analysis here carried out  could be relevant for a large plethora of applications, where multiple reactive sites are concurrently at play. Smart positioning of fully or partially absorbing traps might also translate in innovative non-invasive strategies to control, and consequently shape, the response of a dynamical systems bound to evolve on a complex network-like spatial support.

\section{Materials and Methods}
\subsection{Explicit solution of the 1D Fokker-Planck with two absorbing boundaries}

The solution of the Fokker-Planck equation employed in the main text follows the derivation by \cite{soluzFP}, that we shortly review in the following. Introduce the operator
$F=D\partial^2_x-v\partial_x$ which is defined on $\textsl{D}_F=\{\phi(x)\ |\ \phi(0)=\phi(L)=0\}\cap \textsl{D}^2$ where $\textsl{D}^2$ identifies the set of twice differentiable functions. If
$\phi_{\lambda}\in\textsl{D}_F$ is an eigenfunction of $F$ associated to the eigenvalue $\lambda$, then $e^{\lambda t}\phi_{\lambda}$ is a particular solution of the Fokker-Planck equation (\ref{FP}).  
The operator $F$ can be made Hermitian by defining the scalar product as:

 \begin{equation}
  \langle\phi_1|\phi_2\rangle\equiv\int_0^Le^{-\frac{v}{D}x}\phi_1^{*}(x)\phi_2(x)dx,
  \label{prod_scal}
 \end{equation}

We can therefore find a orthonormal basis formed by the eigenfunctions of $F$ which read:
 \begin{equation}
  \phi_k(x)=\sqrt{\frac{2}{L}}e^{\frac{v}{2D}x}\sin(\frac{k\pi}{L}x)
 \end{equation}
 The associated eigenvalues are $\lambda_k=-\frac{v^2}{4D}-D(\frac{k\pi}{L})^2$.
 \\
 Hence, by denoting with $\phi^0(x)$ the initial probability distribution, one can cast the solution of the Fokker-Planck equation (\ref{FP})  in the explicit form: 
 
 \begin{equation}
  p(x,t)=\sum_{k=1}^{\infty}\langle\phi_k|\phi^0\rangle\phi_k(x)e^{\lambda_k t}.
 \end{equation}
 
 Assuming the initial distribution to be a delta function centered in $x_0=\alpha L$ ($0<\alpha<1$) yields
 \begin{equation}
 \begin{split}
  \langle\phi_k|\phi^0\rangle&=\langle\phi_k|\delta(x-\alpha L)\rangle=\\
&  =\sqrt{\frac{2}{L}}\int_0^Le^{-\frac{v}{D}x}\delta(x-\alpha L)e^{\frac{v}{2D}x}\sin\biggl(\frac{k\pi}{L}x\biggr)=\\
  &=\sqrt{\frac{2}{L}}e^{-\frac{v}{2D}x}\sin\biggl(\frac{k\pi}{L}x\biggr)
  \end{split}
 \end{equation}
 
 from which equation (\ref{p(x,t)}) immediately follows.

\subsection{The case of a generic network: details of the calculation.}

By inserting ansatz  (\ref{espans2}) in equation (\ref{ME}), one gets:

\begin{equation}
 \sum_{\beta}\dot{\hat{p}}_{\beta}(t)u_k^{(\beta)}=\sum_lL^{[i,j]}_{kl}\sum_{\beta}\hat{p}_{\beta}(t)u_l^{(\beta)}.
\end{equation}

To proceed we make explicit the dependence on the eigenvectors:

\begin{equation}
\begin{split}
 \sum_{\beta}\dot{\hat{p}}_{\beta}(t)\sum_{\alpha}C_{\alpha\beta}\psi_k^{(\alpha)}&=\sum_lL^{[i,j]}_{kl}\sum_{\beta}\hat{p}_{\beta}(t)\sum_{\alpha}C_{\alpha\beta}\psi_l^{(\alpha)}=\\
 &=\sum_{\beta}\hat{\rho}_{\beta}(t)\sum_{\alpha}C_{\alpha\beta}\lambda^{(\alpha)}\psi_k^{(\alpha)}
 \end{split}
\end{equation}

Since $\psi^{(\alpha)}$ are linearly independent, one gets: 
\begin{equation}
 \sum_{\beta}\dot{\hat{p}}_{\beta}(t)C_{\alpha\beta}=\sum_{\beta}\hat{p}_{\beta}(t)C_{\alpha\beta}\lambda^{(\alpha)}.
\end{equation}

which yields:

\begin{equation}
 \sum_{\beta}\hat{p}_{\beta}(t)C_{\alpha\beta}=\sum_{\beta}\hat{p}_{\beta}(0)C_{\alpha\beta}e^{\lambda^{(\alpha)}t}.
\end{equation}

Making use of the above relations we obtain:
\begin{equation}
\begin{split}
 p_k(t)&=\sum_{\beta}\hat{p}_{\beta}(t)\sum_{\alpha}C_{\alpha\beta}\psi_k^{(\alpha)}=\\
 &=\sum_{\alpha}\psi_k^{(\alpha)}\sum_{\beta}\hat{p}_{\beta}(0)C_{\alpha\beta}e^{\lambda^{(\alpha)}t}.
 \end{split}
 \label{rho}
\end{equation}
The only quantity that we have to determine is $\hat{p}_{\beta}(0)$, that we wish to express as a function of the initial condition $p_k(0)$. 
To this end we make use of the inverse of (\ref{espans2}):

\begin{equation}
 \hat{p}_{\beta}(t)=\sum_l p_l(t)(u_l^{(\beta)})^*,
 \label{antitrasf2}
\end{equation}
that, introduced into equation (\ref{rho}), results in the general solution reported in the main text. Finally, let us verify that equation (\ref{antitrasf2}) is indeed the inverse of (\ref{espans2}). In formulae:
\begin{eqnarray*}
 p_k(t)&=&\sum_{\beta}\hat p_{\beta}(t)u_k^{(\beta)}=\sum_{\beta}\sum_lp_l(t)(u_l^{(\beta)})^*u_k^{(\beta)} \\
 &=&\sum_l p_l(t)\delta_{kl}=p_k(t)
\end{eqnarray*}
where use has been made of the condition $\sum_{\beta}(u_l^{(\beta)})^*u_k^{(\beta)}=\delta_{kl}$. It is hence clear the importance of dealing with an orthonormal basis to carry out the calculation.

\end{document}